\documentclass[11pt, letterpaper]{article}

\usepackage[colorlinks,urlcolor=blue,linkcolor=blue,citecolor=blue]{hyperref}
\expandafter\def\expandafter\UrlBreaks\expandafter{\UrlBreaks\do\/\do\*\do\-\do\~\do\'\do\"\do\-}
\usepackage{color}
\usepackage{tabularx,booktabs}
\usepackage{url}
\usepackage{fullpage}
\usepackage{graphicx}
\usepackage{lipsum}
\usepackage{cleveref}
 \usepackage{multirow}

\newcommand{\junk}[1]{}

\date{}

\begin{document}


\title{Processing-in-memory for genomics workloads}
\author
{
\centering
William Andrew Simon$^{1}$ \and {Leonid} Yavits$^{2}$ \and Konstantina Koliogeorgi$^{3}$ \and
Yann Falevoz$^{4}$ \and Yoshihiro Shibuya$^{5}$ \and Dominique Lavenier$^{6}$ \and
Irem Boybat$^{1}$ \and Klea Zambaku$^{7}$ \and Berkan Şahin$^{7}$ \and 
Mohammad Sadrosadati$^{3}$ \and Onur Mutlu$^{3}$ \and Abu Sebastian$^{1}$ \and
Rayan Chikhi$^{5}$ \and The BioPIM Consortium \and Can Alkan$^{7}$\footnote{Corresponding Author. Email: calkan@cs.bilkent.edu.tr}
\\
$^{1}$IBM Research Zurich, Switzerland\\
$^{2}$Bar-Ilan University, Tel Aviv, Israel\\
$^{3}$ETH Zurich, Switzerland\\
$^{4}$UPMEM, Grenoble, France\\
$^{5}$Institut Pasteur, Paris, France\\
$^{6}$Université de Rennes, CNRS-IRISA, Inria, Rennes, France\\
$^{7}$Bilkent University, Ankara, Turkey\\
}


\markboth{THEME/FEATURE/DEPARTMENT}{THEME/FEATURE/DEPARTMENT}
\maketitle

\begin{abstract}
Low-cost, high-throughput DNA and RNA sequencing (HTS) data is the main workforce for the life sciences. Genome sequencing is now becoming a part of Predictive, Preventive, Personalized, and Participatory (termed 'P4') medicine. All genomic data are currently processed in energy-hungry computer clusters and centers, necessitating data transfer, consuming substantial energy, and wasting valuable time. Therefore, there is a need for fast, energy-efficient, and cost-efficient technologies that enable genomics research without requiring data centers and cloud platforms. We recently started the BioPIM Project to leverage the emerging processing-in-memory (PIM) technologies to enable energy and cost-efficient analysis of bioinformatics workloads. The BioPIM Project focuses on co-designing algorithms and data structures commonly used in genomics with several PIM architectures for the highest cost, energy, and time savings benefit. 
\end{abstract}


\begin{figure}[htb]
    \centering
    \includegraphics[width=\linewidth]{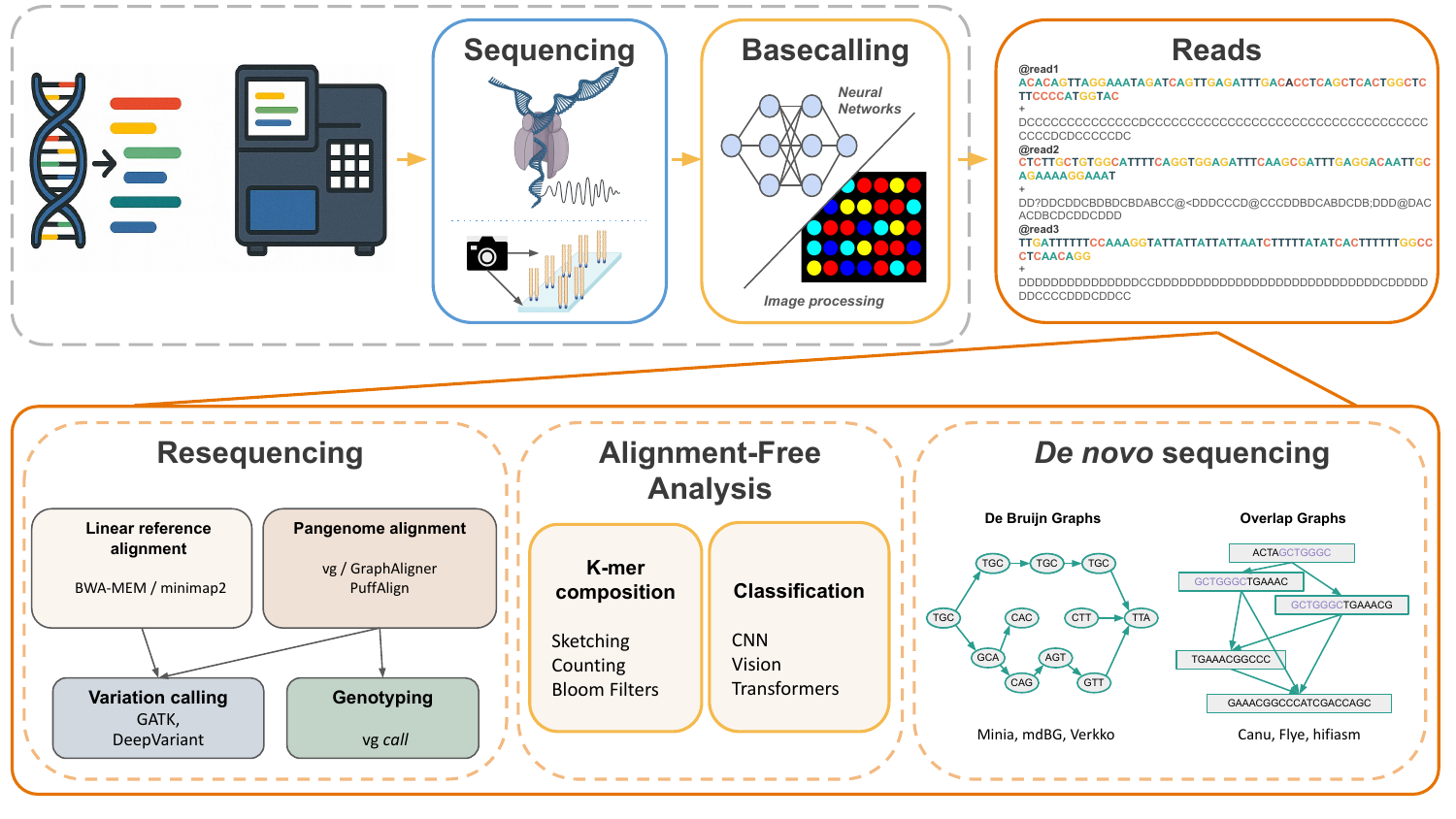}
    \caption{Typical workflow in genomics applications. DNA is extracted and then sequenced using a sequencing instrument, which produces electrical resistance profiles, images, or ``movies''. This signal data is converted into short DNA sequences, called \textit{reads}, during the basecalling step. Reads are then either aligned to an existing reference (resequencing), analyzed independently (alignment-free), or assembled into longer contiguous sequences (\textit{de novo} sequencing).}
    \label{fig:genomics-workflow}
\end{figure}

Genomic data production is expanding exponentially, even in public databases~\cite{Katz2022}. The Illumina NovaSeq X Series can sequence $\sim$20,000 genomes (2400 TB) annually per instrument, while platforms from Pacific Biosciences and Oxford Nanopore Technologies (ONT) continue to improve in both throughput and accuracy. Handheld devices like ONT MinION enable field sequencing for outbreak analysis, and clinical genome sequencing for diagnosis and treatment guidance is increasing. These advances suggest millions of genomes will be sequenced annually. However, rapid analysis remains a major challenge, particularly in portable computing environments for real-time decisions.  

Analysis of genomic data involves many steps, most of which are computationally intensive. Figure~\ref{fig:genomics-workflow} outlines a sequencing-based genomics workflow~\cite{Lappalainen2019}. DNA is extracted and prepared for sequencing. The instrument generates raw data (e.g., images, movies, or electrical resistance profiles), which base calling converts into DNA reads with quality scores. For simplicity, we categorize the sequence analysis into three classes: 1) resequencing when comparing to existing references, 2) alignment-free analysis, and 3) \textit{de novo} assembly. Further analyses include variant effect prediction, genome annotation, and phylogenetics. Similar methods apply to RNA sequencing (\textit{transcriptomics})~\cite{Conesa2016}.

Due to the large size of genomic data and the high computational requirements of algorithms in analyzing such data, several previous studies aimed to accelerate bioinformatics workloads using SIMD, FPGAs, and GPUs, such as SneakySnake~\cite{Alser2019}, Dragen, and ParaBricks, as well as entire analysis pipelines with GPU acceleration (urWGA). 
However, these solutions are still bottlenecked by data movement overhead.
Processing-in-memory (PIM) technologies have emerged as promising approaches that fundamentally address the data movement bottleneck and enable effective parallelization.
Transferring all data to a centralized processor - often located at a considerable distance from the storage system -  leads to latency issues, bandwidth limitations, and high energy demands associated with data transfer and processing. On the other hand, PIM technologies facilitate computation at the data's location. By eliminating the need for extensive data movement, PIM can achieve performance and energy-efficiency improvements by several orders of magnitude when processing large-scale datasets. 

We started the BioPIM Project (\href{https://www.biopim.eu}{https://www.biopim.eu}) to address computational challenges of genomics workloads, leveraging the advantages of PIM computing. BioPIM aims to realize cheap, ultra-fast, ultra-low energy processing that eliminates dependence on large, power-hungry computing clusters/data centers. One of the main goals of BioPIM is to characterize algorithms and data structures commonly used in bioinformatics, including alignment and mapping algorithms, data structures such as Bloom Filters, hash tables, and graphs, and machine learning and deep learning methods. We have found that most data- and compute-intensive applications are memory-bound, therefore good candidates for PIM acceleration (\href{http://doi.org/10.5281/zenodo.14858721}{10.5281/zenodo.14858721}).
BioPIM's research focuses on developing alternative PIM architecture designs for these applications and exploring their potential to accelerate genomic sequence analysis.

Our goal in this paper is to present the ongoing BioPIM efforts on building efficient PIM architectures for genome sequence analysis. 
To this end, we (1) provide the necessary background on state-of-the-art approaches for Processing-in-Memory, i.e., Processing-near-Memory (PnM) and Processing-using-Memory (PuM); (2) introduce PIM architectures developed within BioPIM for genome analysis (Table~\ref{tab:biopim-contributions}), leveraging different techniques for both PnM and PuM; (3)~target fundamental computational kernels in genome analysis as well as different genome analysis stages; and finally (4) discuss future steps and potential impact in bioinformatics research.

\begin{table}[t]
\caption{BioPIM Contributions}
\label{tab:biopim-contributions}
\resizebox{\linewidth}{!}{\begin{tabular}{|c|l|l|l|l|}
\hline
\multicolumn{1}{|l|}{\textbf{Technique}}     & \textbf{Architecture}        & \textbf{Algorithm / Data Structure}                 & \textbf{Repository$^\ast$}                                & \textbf{Citation}      \\ \hline
\multicolumn{1}{|c|}{\multirow{7}{*}{PnM}} & \multicolumn{1}{l|}{2D DRAM} & \multicolumn{1}{l|}{Bloom Filters} & \multicolumn{1}{l|}{PimBloomFilters}        & \multicolumn{1}{c|}{-} \\ \cline{2-5} 
\multicolumn{1}{|c|}{}                     & \multicolumn{1}{l|}{2D DRAM} & \multicolumn{1}{l|}{KSW2}          & \multicolumn{1}{l|}{usecase\_dpu\_alignment} & \multicolumn{1}{c|}{\cite{Mognol2024}}  \\ \cline{2-5} 
\multicolumn{1}{|c|}{}                     & \multicolumn{1}{l|}{2D DRAM} & \multicolumn{1}{l|}{Smith-Waterman-Gotoh}              & \multicolumn{1}{l|}{alignment-in-memory}                              & \multicolumn{1}{c|}{\cite{Diab2023}}  \\ \cline{2-5} 
\multicolumn{1}{|c|}{}                     & \multicolumn{1}{l|}{2D DRAM} & \multicolumn{1}{l|}{WaveFront Algorithm (WFA)}              & \multicolumn{1}{l|}{alignment-in-memory}                              & \multicolumn{1}{c|}{\cite{Diab2023}}  \\ \cline{2-5} 
\multicolumn{1}{|c|}{}                     & \multicolumn{1}{l|}{2D DRAM} & \multicolumn{1}{l|}{Read data set compression}              & \multicolumn{1}{l|}{MiMyCS}                              & \multicolumn{1}{c|}{\cite{Moor2024}}  \\ \cline{2-5} 
\multicolumn{1}{|c|}{}                     & \multicolumn{1}{l|}{2D DRAM} & \multicolumn{1}{l|}{Sorting algorithms}              & \multicolumn{1}{l|}{pim-sort}                              & \multicolumn{1}{c|}{-}  \\ \cline{2-5} 
\multicolumn{1}{|c|}{}                     & \multicolumn{1}{l|}{2D DRAM} & \multicolumn{1}{l|}{GAPiM (Pair-HMM)}              & \multicolumn{1}{l|}{UPMEM\_HaplotypeCaller}                              & \multicolumn{1}{c|}{\cite{Abecassis2023.07.26.550623}}  \\ \cline{1-5} 
\multicolumn{1}{|c|}{\multirow{5}{*}{PuM}}                     & \multicolumn{1}{l|}{Analog PuM} & \multicolumn{1}{l|}{AL-Dorado}              & \multicolumn{1}{c|}{-}                              & \multicolumn{1}{c|}{\cite{Simon2025}}  \\ \cline{2-5} 
\multicolumn{1}{|c|}{}                     & \multicolumn{1}{l|}{SRAM-based CAM}        & \multicolumn{1}{l|}{GCOC}              & \multicolumn{1}{c|}{-}                              & \multicolumn{1}{c|}{~\cite{harary2024gcoc} }  \\ \cline{2-5}
\multicolumn{1}{|c|}{}                     & \multicolumn{1}{l|}{Resistive CAM}        & \multicolumn{1}{l|}{DIPER}              & \multicolumn{1}{c|}{-}                              & \multicolumn{1}{c|}{~\cite{merlin2023diper} }  \\ \cline{2-5}
\multicolumn{1}{|c|}{}                     & \multicolumn{1}{l|}{Gain Cell eDRAM}        & \multicolumn{1}{l|}{k-mer based genome classification}              & \multicolumn{1}{l|}{dash-cam}                              & \multicolumn{1}{c|}{\cite{jahshan2023dash}}  \\ \cline{2-5}
\multicolumn{1}{|c|}{}                     & \multicolumn{1}{l|}{Commodity DRAM}        & \multicolumn{1}{l|}{majority based k-mer matching}              & \multicolumn{1}{c|}{-}                              & \multicolumn{1}{c|}{\cite{jahshan2024majork}}  \\ \cline{1-5}
\end{tabular}}
\vspace*{0.5cm}
\begin{minipage}{\textwidth}
{\footnotesize
    Summary of BioPIM contributions that are outlined in this paper. The full list can be found at \href{https://www.biopim.eu}{https://www.biopim.eu}. $^\ast$Repositories are hosted at \href{https://github.com/BioPIM}{https://github.com/BioPIM}.
    }
\end{minipage}
\end{table}


\section{Background}
\null 
\begin{figure}[htbp]
    \centering
    \includegraphics[width=\textwidth,trim={0cm 2.5cm 0cm 1.5cm},clip]{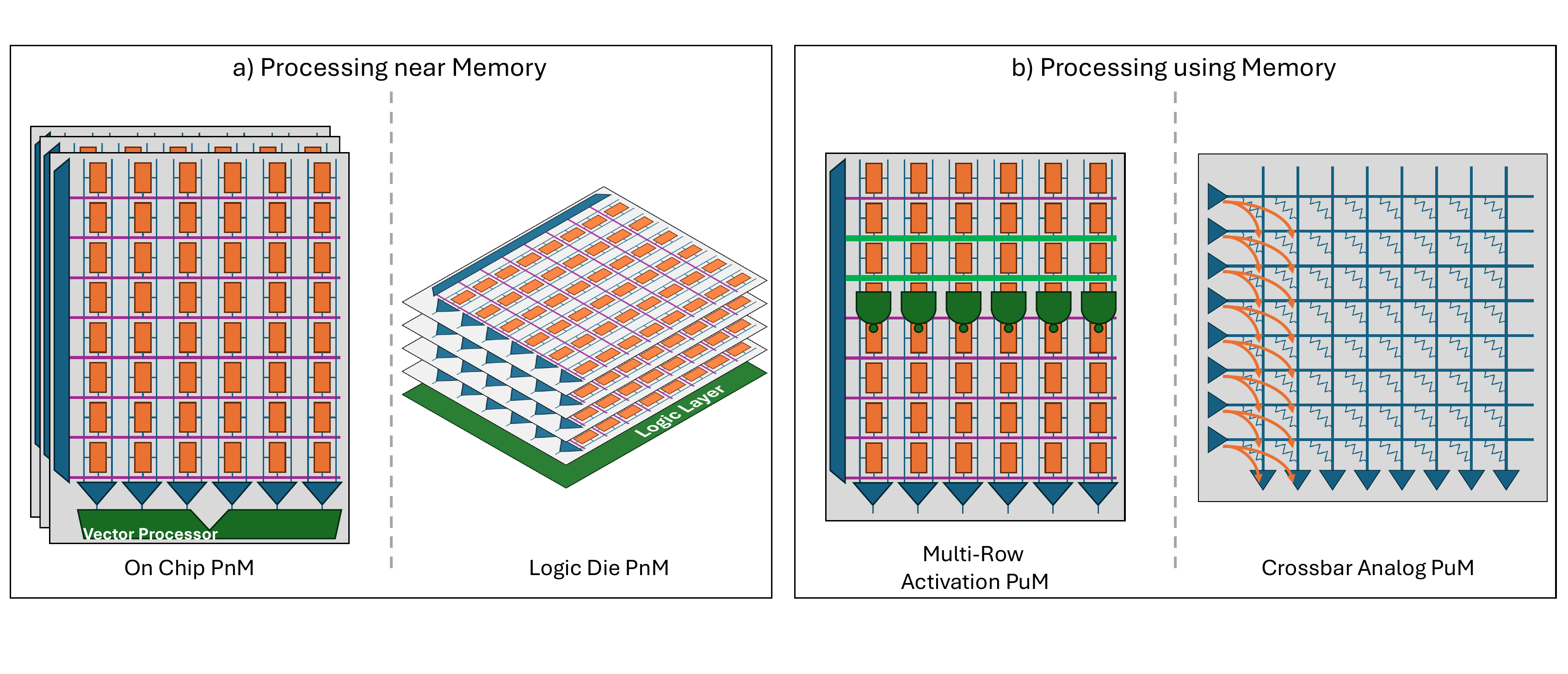}
    \caption{Frameworks of processing a) near and b) in memory. Each method can be accomplished in various ways, with benefits and challenges to each method.}
    \label{fig:pnm-pum}
\end{figure}

\vspace{-10pt} 
There are two main approaches to enabling processing-in-memory in modern systems: (i)~\textit{Processing near memory (PnM)} and (ii)~\textit{Processing using memory (PuM)} (Figure~\ref{fig:pnm-pum}).

The PnM approach adds computation capability to conventional memory controllers, chips, modules, or the logic layer(s) of relatively new 3D-stacked memory technologies.
PnM devices integrate classical memory and logic on the same chip - either on the same die or via a stacked architecture - dramatically reducing data movement. This paradigm shift substantially reduces the energy and latency penalties of traditional systems that rely on moving data between CPUs and DRAM over narrow, high-latency buses.

Emerging PnM solutions include SK Hynix's AiM, Samsung's HBM-PIM, and UPMEM~\cite{Falevoz2024}. Each of these embed processing units close to memory banks to improve performance and energy efficiency. Among these, UPMEM stands out as the only commercially available product at the time of preparing this manuscript. UPMEM integrates lightweight data processing units (DPUs) directly onto the DRAM die without any modifications to the manufacturing process. A typical configuration uses standard DDR4 DIMMs with 16 PIM chips. This architecture delivers thousands of DPUs at the server level, enabling local data processing, consequently dramatically reducing energy costs and data transfer overhead~\cite{Falevoz2024}. UPMEM's robust software stack further simplifies application porting, making it a compelling solution for energy-sensitive, data-intensive workloads such as genomic analysis. However, the current generation of UPMEM DPUs do not support floating-point operations natively; therefore, they require emulated floating-point operations.

On the other hand, PuM uses analog properties of memory arrays (such as Kirchhoff laws or charge sharing) to implement computation within memory arrays directly on the bitlines, reading out the result via sense amplifiers or analog-to-digital converters (ADCs). This approach further alleviates the memory constraints by considerably reducing or sometimes eliminating the need to transfer data outside memory arrays for processing. It also minimizes energy consumption by confining data movements within memory arrays. Moreover, PuM unlocks a very high degree of parallelism by enabling simultaneous processing, for example, in all memory columns in parallel. PnM
places processing units within the memory chip but
outside memory arrays. Hence, it allows for very high
bandwidth (due to a wide access bus) and reduces
energy consumption (due to the close proximity of
processing units to memory). However, PuM improves
energy efficiency significantly beyond that by confining
computing to the memory arrays themselves, within the
frame of the sense amplifiers. A 2-3 orders of magni-
tude energy growth has been reported between the
sense amplifiers’ edge and the CPU. PuM potentially
eliminates this energy growth.  
Key computational workloads, including linear algebra operations and specific genome analyses like k-mer-based classification, exhibit properties well-suited for PuM. These tasks typically feature highly regular, embarrassingly parallel, SIMD-like computations on very large datasets, characteristics that map efficiently onto the inherent parallelism and structural regularity of memory-centric PuM systems.



\clearpage
\section{BioPIM PnM Solutions}

\null

In this section, we present a high-level overview of PnM architectures developed within BioPIM for genomics applications. Our PnM designs implement applications for similarity identification, both alignment-based and alignment-free, and for variant calling. They target real-world PnM systems, i.e., they have been deployed on commercially available UPMEM platforms.

\subsection{Alignment for similarity identification}

Assessment of sequence similarity is a fundamental task in almost all bioinformatics pipelines. Similarities are identified using \textit{alignment algorithms}, which are based on dynamic programming (DP) and therefore have quadratic time and space complexity. In addition to computational challenges, sequence alignment remains a fundamentally
memory-bound computation with a low data reuse ratio. As a result, it suffers from constant data movement latency and a low memory bandwidth bottleneck.
PnM architectures are strong candidates for accelerating alignment, thanks to their ability to reduce data movement overhead and effectively parallelize multiple alignment tasks.
In the following paragraphs, we present a high-level overview of deploying different alignment algorithms on UPMEM. \\

\noindent \textit{KSW2 library.} Using the UPMEM PnM architecture, we have parallelized KSW2, a version of the Needleman-Wunsch algorithm (\href{https://github.com/lh3/ksw2}{https://github.com/lh3/ksw2}). KSW2 is intensively used in Minimap2, one of the most efficient software programs for DNA sequence alignment. 
We distribute sequence batches to each memory unit and perform alignment calculations independently using the computing resources attached to it. We evaluated our design using 9,500 16S RNA sequences and compared it with Minimap2. Our design demonstrates 9x speed-up on an UPMEM server with 160GB PnM memory while achieving $\sim$3.7x reduction in energy consumption compared to a server without PnM memory.\\

\noindent \textit{Smith-Waterman-Gotoh (SWG).} We target the SWG algorithm, a DP-based local alignment method that incorporates an affine gap penalty model. SWG aims to identify similar regions between two sequences without constraints on their location. 
In our implementation, we evenly distribute independent sequence pairs across banks of different DPUs available in the UPMEM platform. A thread is launched in each DPU, and each DPU works independently to align a set of sequence pairs.
We tested our implementation on a UPMEM system with 2560 DPUs (20 UPMEM-DIMMs) running at 425 MHz and compared it with an OpenMP-optimized CPU baseline on server-grade CPU systems. For aligning long-read datasets that strain available memory, our PnM implementation delivers up to 6x speedup over a CPU baseline on an Intel Xeon Gold 5120 server.\\

\noindent \textit{Wavefront Algorithm (WFA).} WFA is an affine gap alignment algorithm that computes exact pairwise alignment efficiently by only using wavefronts in the DP table. This approach significantly reduces memory usage and computational complexity compared to traditional algorithms, making it well-suited for aligning long and highly similar genomic sequences. 
We deploy WFA on UPMEM using the same parallelization and data distribution scheme as in \textit{SWG} and use the same evaluation methodology and setup. Our DPU implementation of WFA outperforms the CPU baselines for small read lengths, exhibiting speedups that range from 1.15x up to 31x.\\

\noindent \textit{CGK embeddings}~\cite{chakraborty_streaming_2016} is a technique for approximating edit distance computations using Hamming distances, which can be calculated in linear time.
Given the randomized nature of the algorithm, averaging multiple independent trials is necessary to reduce errors.
We implemented CGK on the UPMEM PnM architecture, and it is consistently 3x faster than its CPU counterpart in non-trivial situations where more than 2 sequences must be compared. \\

\noindent \textit{Read mapping.} The algorithms outlined above are used to compare pairs of sequences. However, in a genomics experiment, one needs to \textit{map} many reads to a reference sequence. \textit{Read mapping} is a fundamental preprocessing technique with various applications, including variant discovery, genomic data compression, and metagenomic data analysis. Computing the optimal alignments between a reference genome and billions of DNA reads is computationally infeasible. Instead, heuristic approaches are used, leveraging optimal alignment algorithms within a more constrained computational space. The primary challenge lies in efficiently positioning these sequences along the genome while minimizing errors. This task inherently involves massive parallelism, wherein the genome is partitioned across multiple PnM memories, allowing each processing unit to independently identify DNA fragments. In general, localization is facilitated by precomputed indices, which significantly enhance computational efficiency.

Most read mappers use the \textit{seed-filter-extend} strategy, where initial candidate positions are determined by short, exact matches between the target sequence and the reference, which are then extended and optionally filtered iteratively to find the best overall alignment.
The filtering step comes at the cost of reduced sensitivity since alternative alignments can be filtered out to favor the most likely solution.

A competing approach that tries to preserve \textit{all} possibilities is the \emph{seed-embed-extend} in which likely similar regions highlighted by seed matching are embedded into sketched representations.
Edit distance estimations based on sketch comparisons are faster to compute than exact algorithms.
All possible matching regions can then be ranked and reported to the user for a minimal loss of sensitivity. \\

\noindent \textit{Compression.} Read mapping can also help improve data compression efficiency. Using PnM, we implemented a reference-guided algorithm for compressing sequence data. A typical sequencing experiment generates billions of short DNA reads, which are stored in files that are 100s of gigabytes in size. Rather than storing the reads, our algorithm stores their map locations and alignment differences relative to a reference genome. Therefore, the mapping stage is essential, accounting for most of the computational workload in the compression process.  With only 4 UPMEM PnM modules, we achieved 2x speedup and 58\% lower energy consumption than the state-of-the-art DNA read compression tool, Genozip, when run on a server with 16 CPU cores.

\subsection{Alignment-free similarity identification}
\begin{figure}
\centering
\includegraphics[width=\textwidth,trim={1cm 1cm 1cm 1cm},clip]{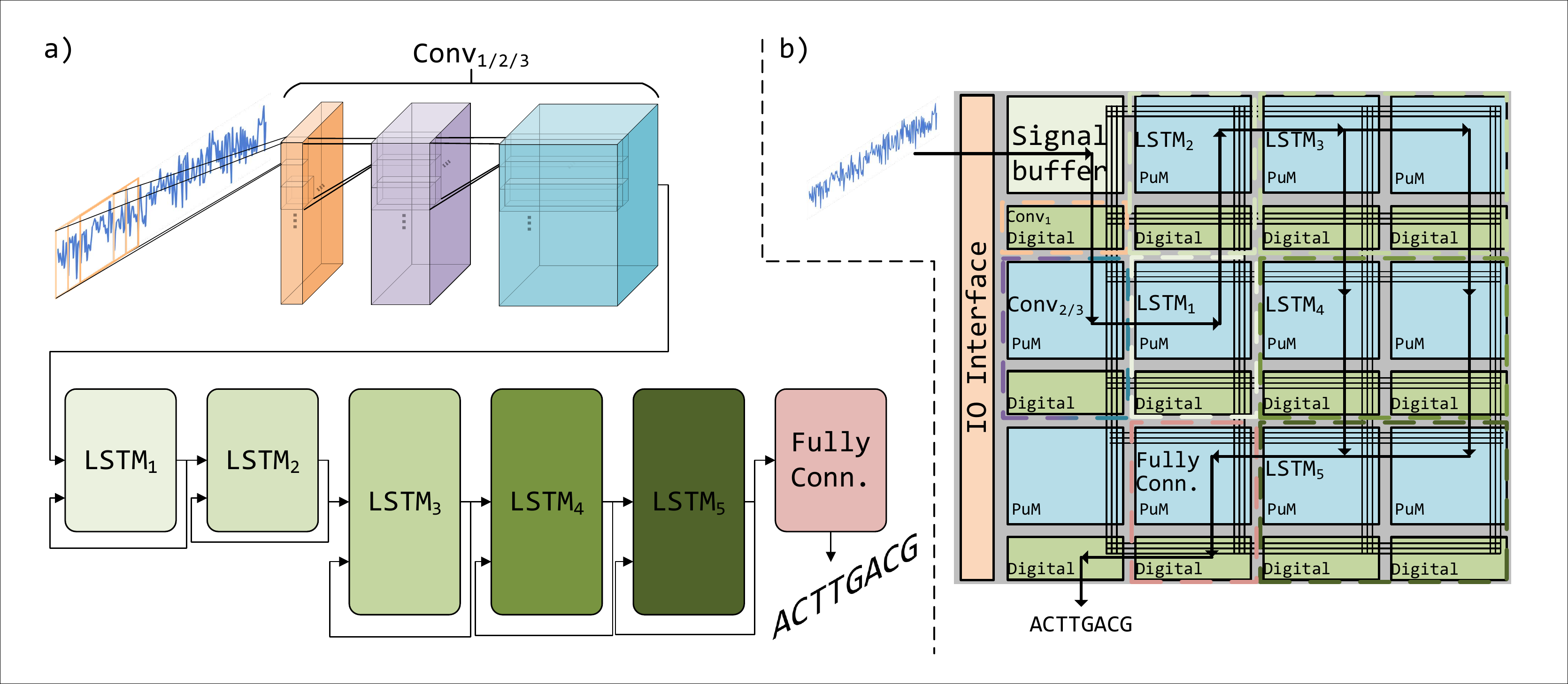}
\caption{The AL-Dorado hybrid CNN-LSTM basecalling network (a) is optimized to map efficiently to the 25mm\textsuperscript{2} CiMBA PuM accelerator (b). Data flows through the accelerator in a pipelined manner, achieving 2x/16.5x better latency/power efficiency against SotA-embedded accelerators.}
\label{fig:CiMBA}
\end{figure}
Similarity searches can also be performed without calculating alignments. This is particularly needed when many genomes need to be analyzed, compared, or classified. These methods are based on the manipulation of \textit{k-mers}, which are short sequences of $k$ nucleotides chosen to represent a genome.  One of the preliminary steps in selecting k-mers is counting them, which typically requires sorting. The enumeration of k-mers in extensive sequencing datasets, as encountered in metagenomics (involving hundreds of billions of k-mers), presents a significant computational challenge. Consequently, substantial research efforts continue to be directed toward optimizing computational time and memory utilization.

We implemented various sorting algorithms on the UPMEM PnM architecture to estimate the potential gain on the general k-mer counting problem. Despite our improvements, the resulting gains are comparatively small compared to those from SIMD acceleration on CPUs with the AVX instruction set.

\subsection{Commonly-used data structures}
Bloom filters are space-saving probabilistic data structures for set membership queries. This data structure is commonly used to test for the presence of k-mers in collections of genomic sequences. Briefly, k-mers are ``inserted'' into a bit-vector using several hash functions, and then any k-mer can be queried in the structure. Bloom filters are also used in de Bruijn graph implementations to boost query efficiency. To improve Bloom filter efficiency and its potential impact across many applications, we have implemented it on the UPMEM PIM architecture. Our evaluation showed $\sim$3x speed up for inserting and querying 100 million elements in a Bloom filter of size 1 GB, compared to a traditional C++ implementation.

Perfect hash functions (PHFs) are a specialized category of hash functions designed to eliminate collisions, ensuring that each unique key maps to a distinct index within the underlying data structure. This property enables constant-time ($O(1)$) lookup operations, significantly enhancing the computational efficiency of applications that rely on fast key retrieval.
A minimal perfect hash function (MPHF) further constrains the output range such that the indices exactly span the size of the input key set.
The minimum space required for such structures is known to be 1.44 bits per key. This efficiency stems from the fact that MPHFs assume a fixed, predefined set of keys, eliminating the need to store the keys themselves within the data structure—a key distinction from conventional hash tables.

PHFs serve as a foundational component in the design of more sophisticated algorithms, including biological data indexing. Over the years, numerous construction algorithms have been proposed, each offering distinct trade-offs regarding space efficiency, build time, and lookup performance.

Among these, \textit{SicHash} distinguishes itself by its highly parallelizable key-placement strategy during construction. The algorithm partitions keys into uniformly sized buckets and subsequently uses cuckoo filters to map each key to its position.

We adapted SicHash's cuckoo filter construction phase for UPMEM's PnM framework to evaluate its performance on alternative computing architectures. However, our preliminary experiments indicate suboptimal performance, with the CPU-based implementation consistently outperforming the PnM variant, even for large key sets.

\subsection{Genomic variation calling}

Variant calling is a critical step in genome analysis, where mutations are identified by comparing the mapping information of reads generated from a sequenced genome to a reference. One of the most computationally intensive parts of this process is the Pair-Hidden Markov Model (Pair-HMM) algorithm, which calculates the probability of each read originating from each potential haplotype. This step has traditionally been a bottleneck on CPU- or FPGA-based platforms due to extensive data movement and heavy floating-point computation. Using the UPMEM PnM framework, these challenges can be overcome by moving the computation directly into memory, significantly reducing data transfer overhead and power consumption. In our approach (GAPiM)~\cite{Abecassis2023.07.26.550623},
we modified the Pair-HMM algorithm to operate in the logarithmic domain and use fixed-point arithmetic to replace slower multiplication and emulated floating-point operations. This innovation not only maintains acceptable levels of accuracy but also delivers up to 2x the speed of CPU SIMD implementations and up to 3x the speed of FPGA solutions. These results demonstrate that even complex, data-intensive tasks such as variant calling can be performed efficiently on PIM platforms, ushering in a new era of scalable, energy-efficient genomic analysis.


\section{BioPIM PuM Solutions}
\null

This section presents an overview of the PuM architectures we developed under the BioPIM project, with a focus on their application in two key bioinformatics tasks: neural network-based basecalling and pathogen classification.

\subsection{Basecalling using in-memory computing crossbars}

As illustrated in Figure \ref{fig:genomics-workflow}, most bioinformatics pipelines begin by sequencing a genomic sample. This involves passing the sample through a flow-cell sensor array and analyzing the DNA strands to infer the nucleotide sequence of each. While various techniques may be applied to perform this analysis, Oxford Nanopore Technology's ultra-long read technology enables sequencing of DNA strands millions of bases long, reducing assembly complexity and improving accuracy on repetitive reads. 

The basecalling step is at the heart of ONT's long-read sequencing workflow. The flow cell produces electrical signals as DNA strands are pulled through nanopores. These electrical signals are processed by a Deep Neural Network (DNN), which infers the original nucleotide sequence. DNNs improved basecalling accuracy by over 10\% compared to previously used hidden Markov models at the time of their introduction in 2017. These DNN models have been improved since then, achieving accuracies of 99.3\% on benchmarks using the latest R10 chemistry. However, this basecalling step is computationally expensive, consuming up to 40\%  of the sequencing workflow and requiring expensive DNN accelerators, typically workstation GPUs, and potentially embedded GPUs.

Using DNNs for basecalling offers an opportunity to accelerate PuM via non-volatile memory-based crossbar arrays. We performed extensive analysis and HW/SW co-optimization of ONT's current basecaller Dorado (also known as Bonito) network for implementation on Phase Change Memory (PCM)-based PuM crossbars. Figure~\ref{fig:CiMBA}-b illustrates a complete PuM architecture for performing on-device basecalling entitled CiMBA (Compute-in-Memory Basecalling Accelerator)~\cite{10924297}, measuring only 25mm\textsuperscript{2} and capable of being co-located at the data collection point. 

The AL-Dorado network, illustrated in Figure~\ref{fig:CiMBA}-a, is, in turn, optimized for implementation on the PuM architecture by adjusting its layer sizes to maximally utilize the available PuM resources and is retrained to be robust against PCM nonidealities and noise sources using the popular AIHWKIT library~\cite{buchel2024aihwkit}. We validate our network retraining accuracy on real PCM-based crossbar arrays, which allows us to quantitatively evaluate the impact of PuM non-idealities. We find that implementing the network in PCM arrays without retraining results in a 7.66\% loss in accuracy, whereas retaining the network to reduce noise reduces this drop to only 1.96\%.

We tested the entire system in a PuM-based simulator to analyze runtime latency, energy, and area performance of DNNs on PuM crossbar architectures. We find that, due to the massive parallelism and pipelining capabilities provided by PuM-based architectures, we can achieve a throughput 2x that of SotA-embedded GPU-based basecalling solutions and 24x the required throughput for real-time basecalling on portable ONT sequencing devices. The architecture is also 16.5x more power-efficient than SotA embedded GPUs, improving field utility by enabling energy-limited setups, such as those powered by batteries, to operate for longer periods.

As basecalling is the first and computationally expensive step in any genomics pipeline, we believe and have demonstrated that performing this step with PuM technologies can benefit all workflows in which it is a precursor, thus accelerating the genomic data analysis.

\subsection{Pathogen genome classification using Content Addressable Memory}
Accurate classification of pathogen genomes from unstructured, sparse sequencing data is essential for disease surveillance, outbreak tracking, and antimicrobial resistance monitoring. Pathogen classification identifies virulent strains, pathogen transmission dynamics, and genetic variations contributing to pathogenicity. With the increasing use of metagenomics, large-scale sequencing datasets require robust computational methods to efficiently classify pathogens. 

Our classification platform uses \textit{k-mer matching}, a heuristic that queries a reference database for newly sequenced DNA fragments. Our unique solution implements pathogen genome classification \textit{in} a special class of memory - a similarity search capable Content Addressable Memory (SAS-CAM).

\begin{figure}[!t] 
     \centering
     \includegraphics[width=1\textwidth]{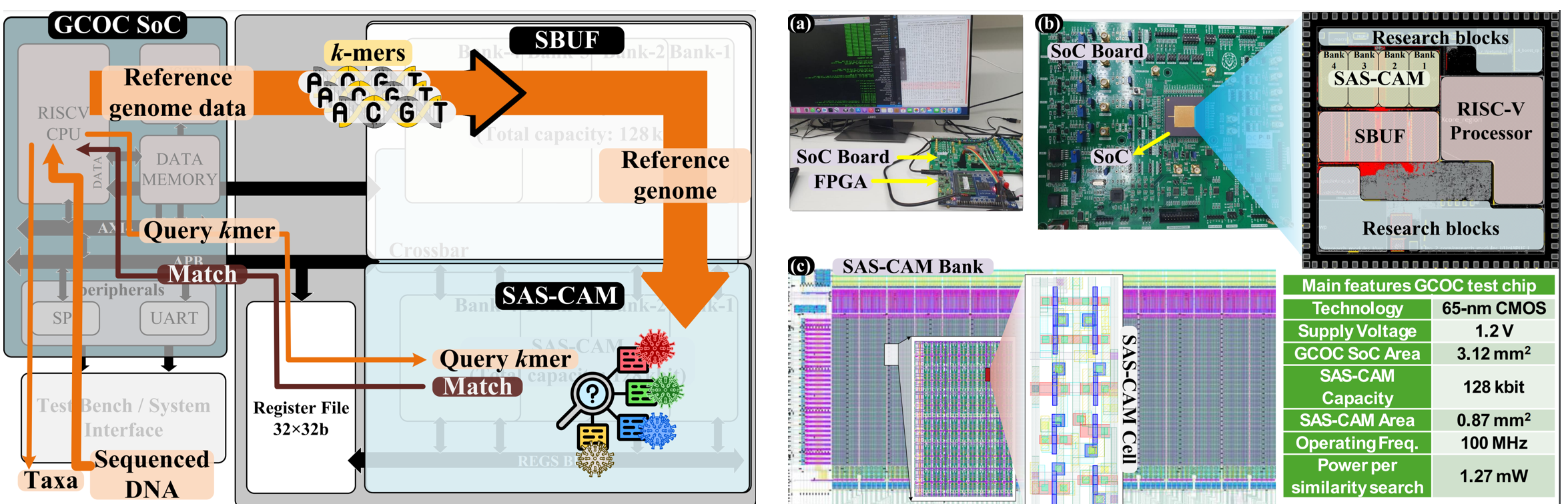} \vspace{-3mm}
     \caption{GCOC genome classifier SoC: (a) GCOC evaluation setup, (b) GCOC SoC test board and layout showing the RISC-V, streaming buffer, and SAS-CAM modules, and (c) SAS-CAM bank and cell layout}\vspace{-5mm}
     \label{fig:gcoc-genome-classifier}
    
\end{figure}

GCOC~\cite{harary2024gcoc}, 
developed within BioPIM, is a fast and highly sensitive approximate k-mer matching-based on-chip genome detection and classification platform.
GCOC SoC architecture, data flow, and evaluation setup are presented in Figure \ref{fig:gcoc-genome-classifier}. 
GCOC comprises SAS-CAM, a dual-issue RISC-V CPU, and a Streaming Buffer (SBUFF), which enables quick uploads and updates to the reference database in SAS-CAM. GCOC SoC has been manufactured using a commercial $65nm$ process. 

SAS-CAM is a 9T NOR CAM with a matchline shared across each row and row sense amplifiers. Its key distinction is an additional NMOS transistor in each CAM cell, which controls the matchline discharge speed. The similarity tolerance threshold is set by adjusting this transistor’s gate voltage and the row sense amplifiers’ reference voltage, enabling dynamic control over classifier sensitivity, specificity, and precision.

Figure \ref{fig:gcoc-genome-classifier} illustrates genome classification in GCOC. The reference genome database (k-mers of predefined organisms) is uploaded offline to SAS-CAM through SBUFF, which RISC-V feeds more slowly. 
During classification, RISC-V fetches newly sequenced DNA reads and extracts k-mers from them as queries for SAS-CAM.
The pointers to the matching k-mers are fed back to RISC-V, which associates them with known taxa. 


DNA sequencing errors and genetic variations, especially in viruses like COVID-19, create edit distance challenges for classifiers. SAS-CAM addresses this by tolerating edit distance mismatches and offering a programmable threshold to adapt to different sequencing error profiles.

GCOC evaluation setup, test board, SoC, and SAS-CAM layout are shown in Figure \ref{fig:gcoc-genome-classifier}. GCOC processing using CAM design enables low-power operation (only 1.27 mW during similarity search) and throughput of up to 400 M k-mers/sec at 100 MHz. Our on-silicon experiments show that GCOC can achieve 96\% sensitivity and 100\% specificity in bacterial and viral genome classification with a Hamming distance threshold of 3. 

While GCOC uses SRAM-based CAM, alternative memory infrastructures exist. DASH-CAM~\cite{jahshan2023dash} 
employs Gain Cell eDRAM for k-mer-based genome classification, expanding the on-chip reference database by 5.5x. DIPER leverages resistive RAM (ReRAM)~\cite{merlin2023diper} 
for greater capacity and support for both Hamming and edit distance tolerances, improving accuracy. Finally, Major-K~\cite{jahshan2024majork} utilizes commodity DRAM for large-scale, cost-effective, and accessible k-mer matching.




\clearpage
\section{CONCLUSION}
\null

The initial promising results of BioPIM architectures, both within PnM and PuM frameworks, reveal that genome analysis can greatly benefit from the PIM approach and provide motivation for an active research field. A critical analysis of the results can provide valuable insights into the requirements and modifications needed for current PIM platforms to support efficient genome analysis. We note that data-centric computing may also be achieved with near-storage-controller~\cite{Ghiasi2024}, in-flash~\cite{Park2022}, or near-router~\cite{Nadig2023} computing, which we did not discuss in this paper.

While UPMEM's current DPUs have proven the technology's ability to speed up and reduce the cost and energy consumption of many applications compared to CPUs, GPUs, and FPGAs, they are limited by DRAM manufacturing constraints. These lightweight processors have limited computational power; for example, floating-point operations are emulated in software rather than implemented directly in hardware.
To address the challenges of modern AI workloads, UPMEM proposed a dedicated PIM-AI chip~\cite{ortega2024pimainovelarchitecturehighefficiency}. This next-generation design uses a stacked architecture and integrates tensor and vector units to meet the needs of deep learning across all domains, including genomics.

CiMBA's energy efficiency and real-time throughput enable on-chip basecalling, facilitating future applications of genome sequencing outside of a lab environment. While these results show great promise on current-generation LSTM-style basecalling neural networks, future basecalling methods are likely to use transformer-based networks. Analyzing and expanding CiMBA's flexibility in accelerating such networks will enable it to serve next-gen sequencing pipelines similarly. The development of basecalling networks specifically for CiMBA, similar to the AL-Dorado network presented here, would enable the best performance and accuracy possible for PuM-accelerated genome sequencing.

GCOC SoC demonstrates the viability of PuM-based genome classification and highlights the potential of integrating k-mer matching and similarity search directly with PuM. These findings support the growing interest in domain-specific hardware for genomic workloads and motivate continued research in PuM-based computing for bioinformatics.
PuM-based pathogen detection and classification offers a wide range of applications, from pandemic preparedness to antimicrobial resistance screening at points of care, environmental bio-surveillance, food safety inspection, and in-field pathogen detection in agriculture.

Community adoption of PIM technologies will require easy programmatic access to these devices for programmers who do not need an in-depth understanding of the underlying architectures. Application programming interfaces (APIs) and extensions to existing programming languages that hide hardware implementation from the programmer have made other accelerators widely used. For example, CUDA and OpenCL, along with APIs such as TensorFlow and PyTorch, are among the main drivers of the immense popularity of GPGPU systems. Therefore, PIM technologies will also require such software layers for both general-purpose and domain-specific use cases. To this end, the BioPIM Consortium is currently developing a bioinformatics domain-specific API (BioPIM Library - BPL), in addition to a general-purpose API (SimplePIM), and an API for efficient computation of transcendental functions (TransPimLib).

This manuscript demonstrates the acceleration of bioinformatic workloads on the UPMEM platform, validating real-world Processing-near-Memory (PnM) capabilities and our proposed PuM designs. While UPMEM represents a specific implementation, the insights gained here are transferable to emerging architectures such as HBM-PIM and CXL-PNM. These platforms share the critical ability to mitigate the data-movement bottleneck caused by high-latency external memory accesses. This architectural shift is essential because algorithmic evolution in genomics—ranging from Smith-Waterman alignment and variation calling to high-throughput base calling and metagenomic analysis—is creating stochastic memory access patterns that defeat standard caching hierarchies. Consequently, the gap left by traditional hardware limitations cannot be filled by simply increasing compute frequency; it requires adopting architectures that prioritize in-situ computation. To fully address these future use cases, research must focus on designing efficient communication frameworks for these commodity PIM devices, effectively decoupling genomic throughput from the limitations of memory bandwidth.

Data production and associated analysis requirements are growing exponentially across all scientific domains. Beyond genomics, fields such as astronomy, high-energy particle physics, digital health, and image recognition are encountering a similar increase in data volume. Current methods for analyzing such large amounts of data require large, power-hungry computing clusters and cloud infrastructure, where substantial energy and time are wasted on data movement. This increased energy demand, in turn, places additional strain on energy production capacities and potentially exacerbates challenges related to resource management. 
Therefore, there is a pressing need to co-develop efficient architectures and algorithms, especially for ``big data'' domains.  
Processing-in-memory technologies, while still being in their infancy, promise new opportunities to accelerate bioinformatics workloads while reducing energy requirements.

\section{ACKNOWLEDGMENTS}

All publications by the BioPIM consortium, including those presented in this manuscript, are available at \href{https://biopim.eu/publications/}{https://biopim.eu/publications/}.
The BioPIM Project is supported by the European Union’s Horizon Programme for Research and Innovation under grant agreement No. 101047160, and the Swiss State Secretariat for Education, Research and Innovation (SERI) (Grant 22.00076).

\def\refname{REFERENCES}

\bibliographystyle{ieeetr}
\bibliography{calkan,biopim}












\end{document}